\documentclass[twocolumn,showpacs,preprintnumbers,aps,prl,10pt,a4paper,superscriptaddress]{revtex4-2}

\usepackage{amssymb}
\usepackage{amsmath}    
\usepackage{graphicx}   
\usepackage{color}      
\usepackage{subfigure}  
\usepackage{float}
\usepackage{bm} 
\usepackage{soul} 
\usepackage{color} 
\usepackage[makeroom]{cancel} 
\usepackage[latin1]{inputenc}
\usepackage{multirow}

\begin{document}

\title{Soft coring: how to get a clarinet out of a flute? }

\author{Frédéric Lechenault}
\email{frederic.lechenault@phys.ens.fr}
\author{Iyad Ramdane}
\author{Sébastien Moulinet}
\affiliation{Laboratoire de Physique de l'Ecole Normale Supérieure, ENS Paris, Université PSL, CNRS, Sorbonne Université, Université Paris Diderot, Sorbonne Paris Cité, 75005 Paris, France.}
\author{Martin Roman-Faure}
\author{Matteo Ciccotti}
\email{matteo.ciccotti@espci.fr}
\affiliation{Sciences et Ingénierie de la Matière Molle, ESPCI Paris, Université PSL, CNRS, Sorbonne Université, 75005 Paris, France.}


\date{\today}

\begin{abstract}
Cutting mozzarella with a dull blade results in poorly shaped slices: the process occurs in a configuration so deformed as to yield unexpectedly curved surfaces. We study the rich morphogenetics arising from such process through the example of coring: when a thin cylindrical hollow punch is pushed into a soft elastomer, the extracted core is "clarinet-shaped", reaching diameters far smaller than those of the tool. With contributions from fracture mechanics and large strain theory, we build a simple yet quantitative understanding of the observed shapes, revealing the crucial role of friction.
\end{abstract}

\maketitle

\begin{figure}[htb]
 \centerline{\includegraphics[width = 0.7\linewidth]{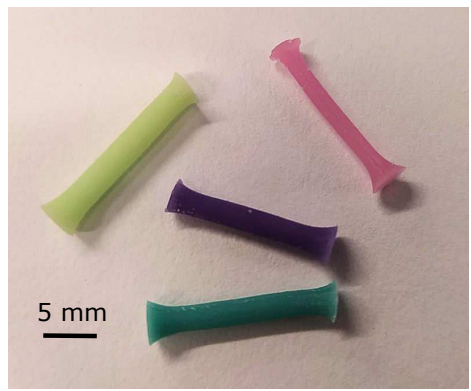}}
 \caption{Some examples of clarinet-shaped cores obtained by cutting rubber with a cylindrical punch. In this case, the punch had a radius of 2mm, as witnessed by the similar size of the core extremities, but the core radius changes as the samples differ in stiffness.}
  \label{fig:Clarinets}
\end{figure}

When we use a cookie cutter on a thin layer of dough we obtain the shape we meant, e.g.\ a gingerbread man! Yet when coring soft elastic materials like rubber with long cylindrical punches, we are surprised to obtain thin clarinet shapes like those shown in Fig.\ \ref{fig:Clarinets}.
As a matter of fact, while both ends of the core have the same radius $R$ as the coring blade, the central part of the core reaches a steady-state radius $R_{core}$ which can be significantly smaller than the former if the cored substrate is thick enough. The overall shape is surprisingly symmetric despite the fact that the coring process has a well defined direction involving large asymmetric strains.

In most cutting operations, we impose the path of the blade (at least we try to). However, as illustrated by our soft coring experiments, the final shape of the cut parts is selected by the deformation of the sample induced by the blade. In practical situations, in order to enforce a close match between the blade path and the final shape, different strategies are used to minimize this deformation, or to reduce the applied forces. Several lines of work have investigated classical ways to cut efficiently, such as sharpening the blades~\cite{MCCARTHY2007,HUTCHENS2021}, introducing slicing motion~\cite{REYSSAT2012,ATKINS2016,MORA2020,HUI2021}, or minimizing friction~\cite{LAKE1978,GENT1994,HUTCHENS2019}. Notably, a plethora of tools have been developped historically, specialized for each material, cutting speed, and target chip size~\cite{ATKINS2009}.
However, when addressing soft materials such as food, gels or living tissues, these strategies can fail and shape control is lost. 

The soft coring is investigated here as a robust model experiment 
for rationalizing the geometrical and material nonlinearities induced by the coupling of large strains and cutting.
In this situation, the cutting edge of the blade moves into the material after the latter has been compressed by the applied force required for cutting, resulting into a biaxial extension in a plane perpendicular to the direction of cutting. This is the key geometrical nonlinearity of the problem: the final shape of the core results from its elastic recovery once the coring is completed.

\begin{figure}[htb]
 \centerline{\includegraphics[width = 0.6\linewidth]{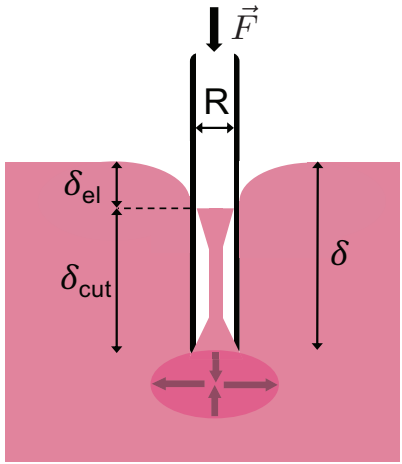}}
 \caption{Sketch of the coring mechanism. The compressed region below the blade of radius $R$ undergoes a contraction of $\lambda_z<1$ in the vertical direction and an equibiaxial expansion of $\lambda_R$ in the horizontal plane. While the external portion of the cut sample remains stretched around the blade, the inner part relaxes into the final core shape. The vertical displacement $\delta$ of the blade can be split into the cut length $\delta_{cut}$ and an elastic component $\delta_{el}$.}
  \label{fig:SKETCH}
\end{figure}

A preliminary insight into the scaling of this phenomenon can be obtained by considering the brittle limit, where deformations are small, achieved when using very sharp tools.
We consider a cylindrical coring tool of radius $R$ pushed with a force $F$ orthogonally into a soft material with Young's modulus $E$, as illustrated in Fig.\ \ref{fig:SKETCH}. For small strain, the cut length $\delta_{cut}$ can be identified with the vertical blade displacement $\delta$. Invoking Saint-Venant's principle we can expect a steady-state cutting regime when the distance of the cutting edge from the external surfaces is larger than $R$, resulting in a core of constant final radius $R_{core} < R$. While this provides a rationale for the clarinet shape of the two edges of the core, the focus here will be on the modeling of the ratio $\lambda_R = R/R_{core}$ for steady-state cutting.
When initially neglecting friction on the blade and bulk energy dissipation, we can describe the cutting process with fracture mechanics tools such as the Griffith energy release rate. The work performed when the blade is advanced by a distance $d\delta$ is:
$$dW = F d\delta \simeq F d\delta_{cut}$$
%
%

The average compressive stress at the scale of the blade radius $R$ is $\sigma \sim {F / R^2}$. The elastic energy $U_{el}$ stored in the compressed sample can be written as
$$U_{el} \sim {\sigma^2 \over E} R^3 \sim {F^2 \over E R}$$
%
which does not depend on the crack length $\delta_{cut}$ in the considered steady-state regime.  
The cutting criterion thus reads
\begin{equation}
G = {dW \over dA} - {\partial U_{el} \over \partial A}= {F \over 2 \pi R} = \Gamma_{cut}
\label{eq:Griffith}
\end{equation}
\begin{equation}
F_{cut} = {2 \pi R \Gamma_{cut}}
\label{eq:Fcut}
\end{equation}
where the cutting energy $\Gamma_{cut}$ represents the cost for cutting a new unit area of material $dA = 2 \pi R d\delta_{cut}$.
The average vertical contraction at the scale $R$ below the blade is thus:
$$\varepsilon_{cut} \sim -{F_{cut} \over R^2 E} \sim -{\Gamma_{cut} \over ER}$$

Due to incompressibility of rubber, this results into a local equibiaxial extension in the horizontal directions $\varepsilon_R = -\varepsilon_{cut}/2$, which sets the scaling for the final radius $R_{core}$ of the cored sample:
\begin{equation}
\varepsilon_R = {R \over R_{core}} -1 \sim { {\Gamma_{cut} \over E R}}
\label{eq:LinearMod}
\end{equation}

The outcome of this preliminary scaling analysis is that the shape of the core region is ruled by the ratio between the size $R$ of the cylindrical cutting blade and a characteristic length scale of the material $\ell_{TE} \equiv \Gamma_{cut}/E$, that we will name ``tomo-elastic'' length scale (from greek $\tau o \mu \acute{o} \varsigma$ = cut). 
%
The morphogenetic effect should only be substantial when $R \leq \ell_{TE}$. A rough estimation of $\ell_{TE}$ can be obtained for rubbers by considering a typical tearing energy of $\Gamma \sim 1000$ J/m$^2$~\cite{LAKE1978}, and modulus $E \sim 1$ MPa, which yields millimetric scales. 

\begin{figure}[htb]
 \centerline{\includegraphics[width = 0.99\linewidth]{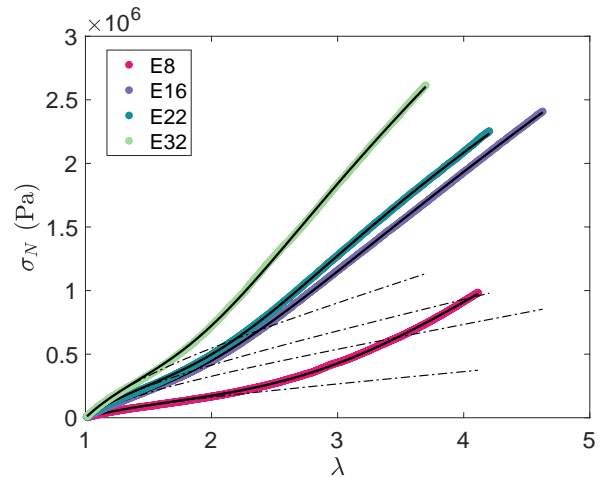}}
 \caption{Uniaxial extension curves (nominal stress against stretch) for the four elastomers ($\bullet$), and the corresponding fit with Eq.~(\ref{eq:ConstModelStress}) (lines), as well as with Neo-Hookean law (dash-dot).}
  \label{fig:FitTraction}
\end{figure}

Following this estimation, we performed systematic experiments by changing the elastic properties of the rubbers and the radius of the coring tools in this regime. 
As model soft elastic rubbers, we selected a family of four commercial silicon-based elastomers (Elite Double\textsuperscript{\tiny\textregistered}, Zhermack). 
Their mechanical properties were measured in a standard uniaxial tensile test and reported in Fig.~\ref{fig:FitTraction}. The corresponding elastic moduli are reported in Table~\ref{tab:table1}.
The samples were cast into 
cylinders 100 mm in diameter and variable heights up to $30$ mm in order to be at least four times larger than the blade radius.
The cylindrical blades were made from a set of steel tubes and syringes, with 15 different diameters ranging from $0.5$ mm to $8$ mm. The blade thickness $h$ remains in approximately constant proportion $R/5$.
The tubes were sharpened using a conical milling tool providing variable internal blade angles $\theta$ from 10$^\circ$ (sharp) to 90$^\circ$ (flat).  

\begin{table}[ht]
\begin{center}
\begin{ruledtabular}
\begin{tabular}{ccccccc}
 Shore A & $E$  & a & b & c & $\Gamma_{cut}$ & $\ell_{TE}$  \\ [0.5ex]
         & (kPa) &   &  &   & (J/m$^2$) & (mm) \\ [0.5ex]
 \hline
 8  & 276 & -0.77& 0.0064 & 1.88 & 1190 & 4.3 \\
 16 & 559 & -0.69& 0.079& 1.35 & 1400 & 2.5 \\
 22 & 709 &  -0.60& 0.046& 1.80 & 1630 & 2.3 \\
 32 & 938 & -0.61& 0.082& 1.66 & 1600 & 1.7 \\ 
\end{tabular}
\end{ruledtabular}
\end{center}
\caption{Material properties of the elastomers. Shore A hardness as provided by the manufacturer. $E,a,b,c$ are the material's parameters obtained by fitting the uniaxial curves in Fig.~\ref{fig:FitTraction} with Eq.~(\ref{eq:ConstModelStress}). $\Gamma_{cut}$ and $\ell_{TE}$ are the values of the cutting energy and tomo-elastic length obtained by fitting the coring data in Fig.~\ref{fig:FitCoring} with version $\alpha$ of Eq.~(\ref{eq:MainModel}).}
\label{tab:table1}

\end{table}



\begin{figure}[htb]
 \centerline{\includegraphics[width = 0.99\linewidth]{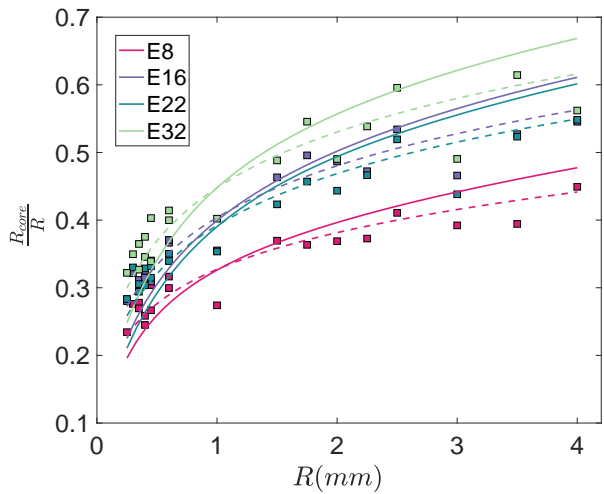}}
 \caption{Measured ($\square$) coring ratios against the radius of the cylindrical blade. The full and dashed lines represent the predictions of models $\alpha$ and $\beta$ respectively (cf.\ Eq.\ (\ref{eq:MainModel}))}.
  \label{fig:FitCoring}
\end{figure}

\begin{figure}[htb]
 \centerline{\includegraphics[width = 0.99\linewidth]{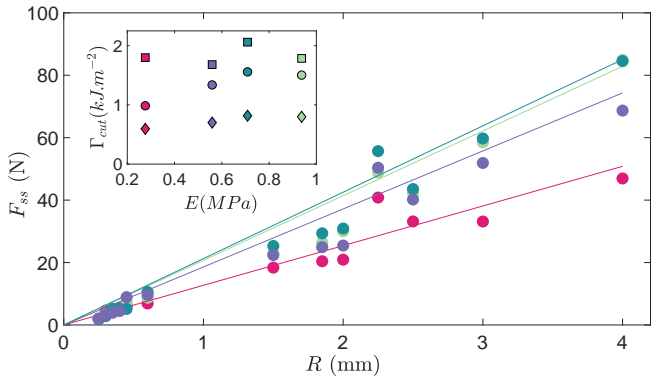}}
 \caption{Scaling of the measured cutting force $F_{ss}$ in the initial stage of steady-state cutting against the blade radius $R$. The inset reports the cutting energy $\Gamma_{cut}$ estimated by force measurements ($\circ$) and by fitting the core shape data from Fig.\ \ref{fig:FitCoring} with model $\alpha$ ($\diamond$) and $\beta$ ($\square$) defined by Eq.~(\ref{eq:MainModel})}.
  \label{fig:peakforce}
\end{figure}

The coring protocol is as follows. A tube is mounted vertically in the jaws of a mandrel attached to the moving part of a testing machine Instron\textsuperscript{\tiny\textregistered} (model 5965) with a 100 N load cell. An elastomer sample is laid on a flat wooden sheet sitting horizontally underneath the tool. The sample is then cored by moving the tube into the elastomer at a velocity of $10$ mm/min until it touches the wooden plate, resulting in a sharp increase in the measured force. The steady-state diameter of the cut part is then measured by taking a high resolution picture with a D800 Nikon camera. 
The experimental results for the core radius $R_{core}$ and cutting force $F_{cut}$ measured in the steady-state regime are reported in Fig.\ \ref{fig:FitCoring} and Fig.\ \ref{fig:peakforce} as a function of the tool radius $R$. The values of $h$ and $\theta$ are not reported here since they had no appreciable influence on the results.


As predicted by the preliminary linear scaling model, the coring experiments confirm that the samples undergo large strains 
when the cylindrical blade radius $R$ becomes comparable or smaller than the millimetric scale estimated for the tomo-elastic length $\ell_{TE}$. Unsurprisingly, the linear model fails to capture the functional dependency of the core size on the blade radius.  Since the observed strains can become larger than 300\%, we developed a more advanced model, based on hyperelasicity, in order to account for both finite strains and nonlinear material behavior. 

From a geometrical point of view, the elastic indentation $\delta_{el}$ becomes larger than the blade radius $R$, and it can no longer be neglected in front of the length $\delta_{cut}$ of the cut material, so that the blade displacement should be written as $\delta = \delta_{el} + \delta_{cut}$ (see Fig.~\ref{fig:SKETCH}).
%
%
The final core radius $R_{core}$ is selected by the value of the in-plane equibiaxial stretch $\lambda_R \geq 1$ of the elastomer in the compressed region of size $R$ just below the cylindrical blade tip. The compressive force $F_{cut} = \pi R^2 \sigma_T$ applied by the blade on the elastomer should be related to the local average true stress $\sigma_T$ by considering the homogeneous uniaxial compression ($\lambda_z \leq 1$) of an incompressible hyperelastic material with energy density $\mathcal{W}(I_1)$, where $I_1$ is the first invariant of the Cauchy strain \cite{OGDEN1985}:
$$I_1 = {\lambda_z^2+{2 \over \lambda_z}} \qquad \lambda_z = {1 \over \lambda_R^2}$$
\begin{equation}
\sigma_N = {d \mathcal{W}(I_1) \over d \lambda_z}
= 2 {d \mathcal{W} \over d I_1} \left( \lambda_z - {1 \over \lambda_z^2} \right)
\label{eq:sigmaNcomp}
\end{equation}
\begin{equation}
\sigma_T = \lambda_z \sigma_N 
= 2 {\partial \mathcal{W} \over \partial I_1} \left( {1 \over \lambda_R^4} - {\lambda_R^2} \right)
\label{eq:sigmaTcomp}
\end{equation}

The value of the local compressive force at the scale $R$ of the blade front during steady-state cutting can also be related to the cutting energy $\Gamma_{cut}$ by a Griffith-like energy balance similar to Eq.~(\ref{eq:Griffith}). %
%
%
During steady-state cutting all the local parameters such as $F_{cut}$ and $\lambda_R$ are constant. In the absence of friction, the elastic component of the displacement $\delta_{el}$ is also constant, and the displacement of the loading point is thus simply related to the variation of the cut length $\delta_{cut}$ as measured in the reference frame: $d\delta = d\delta_{cut}$.
However, the evaluation of the relevant area for the expression of the surface energy term is delicate under large strain: while the blade front effectively advances along a cylindrical surface of radius $R$, the surface cut in the unstrained material is a cylinder of smaller radius $R_{core}$. We chose to initially preserve both choices in the formulation of the model, before discussing their significance and consistency with our measurements:
$$dA = 2 \pi \left\{ R; R_{core} \right\} d \delta_{cut} = 2 \pi {R \over \left\{ 1;\lambda_R \right\}} d \delta_{cut}$$
where the term in brackets $\left\{\alpha;\beta\right\}$ means the alternative between the two different model hypothesis. The Griffith-like cutting condition becomes then:
\begin{equation}
G = {F_{cut} \left\{ 1;\lambda_R \right\} \over 2 \pi R} - {\left\{ 1;\lambda_R \right\} \over 2 \pi R}{d U_{el} \over d \delta_{cut}} = \Gamma_{cut}
\label{eq:G}
\end{equation}

Although the local strain field has a complex shape, when considering the steady-state cutting regime away from external surfaces, the variation of the total elastic energy for a unit cut advance can be estimated by the residual stretch energy in the outer portion of the cut cylinder (the inner part being relaxed as sketched in Fig.~\ref{fig:SKETCH}).
%
For the plane-strain stretch ($\lambda_z =1$) of a cylindrical cavity of radius $R_{core}$ in an incompressible hyperelastic material, the stretch field is uniquely determined by the stretch $\lambda_R$ imposed by the punch as \cite{BIGGINS2019}:
$$\lambda_{r}(r) = {1 \over \lambda_\theta} = 
{1 \over \sqrt{1 + {R_{core}^2 \over r^2}(\lambda_R^2-1)}}$$
$$I_1(r) = 
{1 + {R_{core}^2 \over r^2}(\lambda_R^2-1)} + {1 \over 1 + {R_{core}^2 \over r^2}(\lambda_R^2-1)} + 1$$
$${dU_{el} \over d\delta_{cut}}(\lambda_R) = \int_{R_{core}}^\infty \mathcal{W}(r) 2 \pi r dr$$

According to Eq.~(\ref{eq:G}) the steady-state cutting force $F_{cut}$ can be expressed as:
\begin{equation}
F_{cut} = {\Gamma_{cut} 2 \pi R \over \left\{ 1;\lambda_R \right\}} + {d U_{el} \over d \delta_{cut}}(\lambda_R)
\label{eq:Fblade}
\end{equation}

The value of the radial stretch $\lambda_R$ can thus be determined as a function of the cutting energy by equating the two expressions for the true stress $\sigma_T = -{F_{cut} / \pi R^2}$ between Eq.~(\ref{eq:sigmaTcomp}) and (\ref{eq:Fblade}):
\begin{equation}
{2 \over E} {d \mathcal{W} \over d I_1} \left( {1 \over \lambda_R^4} + {\lambda_R^2} \right) + {1 \over \pi R^2 E}{dU_{el} \over d \delta_{cut}}(\lambda_R) + {2\Gamma_{cut} \over E R \left\{ 1;\lambda_R \right\}} = 0
\label{eq:MainModel}
\end{equation}
where we chose to divide every term by the elastic modulus $E$ in order to obtain a dimensionless equation for $\lambda_R$.
This constitutes the main relationship of our finite strain modeling for the shape of the central steady-state portion of the core.

When comparing with the linear model, we remark that the large strain model contains four different sources of non-linearity, which can be attributed either to large geometry changes or to the large strain material response. The equibiaxial expansion of the compressed region underneath the blade of size $R$ has three different consequences: (1) the hyperelastic uniaxial compression law becomes non-linear due the stretch of the application area (left term); (2) the external part of the cut material remains stretched around the blade and stores elastic energy (middle term); (3) the area of the unstrained cut surface is much smaller than the area swept by the blade (right term). Moreover, the left and middle terms are also affected by the specific nature of the large strain material response encoded in $\mathcal{W}(I_1)$. The competition between these different terms selects a single value of the stretch $\lambda_R$ as a function of the cutting energy $\Gamma_{cut}$, the elastic modulus $E$ and the tool radius $R$. Since the left and middle terms scale as $\mathcal{W}$, which scales with $E$, dimensional analysis confirms that the dimensionless equilibrium stretch $\lambda_R$ must be written as a function of the dimensionless combination $\Gamma_{cut}/E R = \ell_{TE}/R$.

In order to gain more insight into the relative impact of the four nonlinearities, let us first focus on the geometry by restricting our model to a Neo-Hookean energy density functional \cite{BIGGINS2019}:
$$\mathcal{W} = {E \over 6}(I_1 - 3)$$
\begin{equation}
{dU_{el} \over d\delta_{cut}}(\lambda_R) = \pi {E \over 3} R_{core}^2 (\lambda_R^2 - 1) \log \lambda_R
\label{eq:CylindNH}
\end{equation}
\begin{equation}
\left( {1 \over \lambda_R^4} - {\lambda_R^2} \right) + \left(1 - {1 \over \lambda_R^2}\right) \log \lambda_R + 6 {\Gamma_{cut} \over E R \left\{ 1;\lambda_R \right\}} = 0
\label{eq:LambdaNH}
\end{equation}

The solution of this simplified model does not correctly predict the dependency of our experimental data for $\lambda_R$ on the radius of the cutting tool, which can only be captured by taking into consideration the full nonlinear material behavior. However, this Neo-Hookean expression allows to appreciate that the middle term (residual cylindrical stretch) becomes very small in front of the left term (uniaxial compression) when the radial stretch $\lambda_R$ is increased for small values of $R$. 
Moreover, it is easily shown that when expanding any elastic energy potential $\mathcal{W}(I_1)$ into a series of power law terms $(I_1-3)^n$ in order to capture material strain hardening, the contribution to the left term will always remain $n\lambda_R^2$ times larger than the middle term. We can thus comfortably approximate Eq.~(\ref{eq:MainModel}) by ignoring the middle term when evaluating our model for an arbitrary nonlinear constitutive law.

In particular, in order to capture the strain hardening behavior of our elastomers, displayed in Fig.~\ref{fig:FitTraction}, which present an initial softening behavior followed by a linear strain hardening of the $\sigma_N(\lambda)$ curve, we use Eq.~(\ref{eq:sigmaNcomp}) with the following custom expression:
\begin{equation}
{d\mathcal{W}\over dI_1} = 
{E \over 6(1+a)} \left( 1 + {a \over 1 + b(I_1-3)^c} \right)
\label{eq:ConstModelStress}
\end{equation}
%

The fitted parameters for each elastomer are reported in Table~\ref{tab:table1}.
The corresponding solutions of equation (\ref{eq:MainModel}) (where we dropped the middle term) are plotted in Fig.~\ref{fig:FitCoring}. Both versions $\alpha$ and $\beta$ of the model can adjust the experimental dataset very well for each elastomer over a wide range of cutting blade radii. However, the values of the tomo-elastic length $\ell_{TE}^{\alpha,\beta} = \Gamma_{cut}^{\alpha,\beta}/E$ that are extracted by the fitting procedure depend on the choice of the model $\alpha$ or $\beta$. The corresponding values of $\Gamma_{cut}^{\alpha,\beta}$ are reported in the inset of Fig.~\ref{fig:peakforce}.
%

Let us now discuss more extensively the important role of friction in the coring process. When pushing a blade into a soft material we can soundly separate two main contributions of friction, to be respectively associated to the contact of the cut material with the side blade walls and the tip region. The first contribution $F_f$ increases with the progression of the cut length $\delta_{cut}$. In contrast, the second one is localized at the scale of the blade tapered region, which 
in steady-state cutting conditions is expected to be invariant during the advance of the blade.
We remark that only the latter contributes to the local compressive force $F_{cut}$, which is responsible for the selection of the lateral stretch $\lambda_R$ at the moment of cutting, according to the model defined above. This explains why the observed core radius is almost constant during the cutting process, while the total applied force $F = F_{cut} + F_{f}$ is slowly increasing due to increase of the sliding surface.
%
Following the approach of Irwin \cite{IRWIN1957}, the local dissipation energy due to friction on the tapered region can be added to the fracture energy required to break the polymer network resulting into a global energy cost for cutting a unit area of material $\Gamma_{cut}$. However, when considering soft materials, the fracture energy term should be evaluated in the unstrained reference (radius $R_{core}$), while the local friction energy term should be determined by the actual reference (radius $R$), as described respectively by models $\beta$ and $\alpha$ introduced above.
Although the two contributions to $\Gamma_{cut}$ were estimated to have the same order of magnitude \cite{LAKE1978,HUTCHENS2021,MCCARTHY2007}, the relevance of the friction contribution to Eq.~(\ref{eq:MainModel}) will be dominant for increasing values of $\lambda_R$. This explains why only model $\alpha$ provides values of $F_{cut}$ which are physically consistent, i.e.\ lower than the forces $F_{ss}$ measured independently during the initial stages of steady-state cutting, as reported in Fig.~\ref{fig:peakforce} along with the corresponding values of $\Gamma_{cut}$ in the inset. 
This interpretation is also consistent with the relatively weak influence of the blade sharpness on the measured radius $R_{core}$ of the cored regions. Future investigations might allow independant identification of these two contributions.

The presented model captures the main features of the steady-state core radius, the selection of which is mainly dominated by the cylindrical blade radius $R$, the cutting energy $\Gamma_{cut}$ and the elastic modulus $E$, in the form of the ratio between $R$ and the tomo-elastic length $\ell_{TE}=\Gamma_{cut}/E$.
The core shape is thus determined by a combination between geometrical and material nonlinearities. Although the physical phenomenon can be qualitatively explained by the geometrical nonlinearity associated to cutting the soft material in a stretched condition, the quantitative relation between the final core radius and the cylindrical blade radius is mainly determined by the details of the nonlinear - strain hardening - material behavior. We remark that if the material were Neo-Hookean, the geometrical effect would be considerably larger than observed in our experiments, since strain hardening has the effect of limiting the stretch in the compressed region.
Although the role of friction on the side walls of the blade does not affect the shape of the core, local friction on the blade edge has a dominant effect on the value of the cutting energy, and thus on the tomo-elastic length, which cannot be identified with the elasto-cohesive length $\Gamma_{fract}/E$ defined for fracture phenomena \cite{CRETON2016}.
As a final remark, friction is also very important for the origin of the two clarinet shapes in the edges, since it prevents the lateral stretching and thus enforces the core and tool radius to be the same. For the first (incoming) clarinet the lateral stretch is suppressed by static friction of the blade before the initiation of cutting. For the second (outcoming) clarinet the lateral stretch is suppressed by static friction between the sample and the bottom support plate. In both cases, the transient regions shaping the clarinet bells extend over a length R from Saint-Venant's principle.
Our study opens perspectives into complex programmable surface shaping, in a mundane, accessible setting fitted for an array of design applications.

\bibliography{Coring}

\end{document}